\documentclass[useAMS,usenatbib]{mn2e}

\usepackage{graphicx}
\usepackage{subfigure}
 \usepackage{graphicx,amsmath,amssymb}
\usepackage{color} 
\usepackage{hyperref}
\usepackage{color}
\usepackage{amsmath}
\def\lsim{\lower.5ex\hbox{$\; \buildrel < \over \sim \;$}}
\def\gsim{\lower.5ex\hbox{$\; \buildrel > \over \sim \;$}}
\def\simeq{\lower.3ex\hbox{$\; \buildrel \sim \over - \;$}}
\def\ch{\lower-0.55ex\hbox{--}\kern-0.55em{\lower0.15ex\hbox{$h$}}}
\def\lh{\lower-0.55ex\hbox{--}\kern-0.55em{\lower0.15ex\hbox{$\lambda$}}}
\def\eg{{\it e.g.,} }
\def\etal{{\em et al.} }

\def\eg{{\it e.g.,}}
\def\etal{{\em et al.}}

\def\lsim{\lower.5ex\hbox{$\; \buildrel < \over \sim \;$}}
\def\gsim{\lower.5ex\hbox{$\; \buildrel > \over \sim \;$}}
\def \simeq{\lower.3ex\hbox{$\; \buildrel \sim \over - \;$}}

\def\apj{{ApJ}}%
\def\apjl{{ApJL}}%
\def\aap{{A\&A}}%
\def\ijmpd{{IJMPD}}%
\def\mnras{{MNRAS}}%
\def\na{{New Astron.}}%
\def\pasj{{PASJ}}%
\def\MmSAI{{Mem. S.A.It}}

\title[magnetized dissipative accretion flow]
{Dynamical structure of magnetized dissipative accretion flow around black holes}
\author[Biplob Sarkar, Santabrata Das]
{Biplob Sarkar\thanks{E-mail:
biplob@iitg.ernet.in (BS); sbdas@iitg.ernet.in (SD)}, 
Santabrata Das\footnotemark[1] \\ 
$$Indian Institute of Technology Guwahati, Guwahati, 781039, India.
}
\begin{document}

\date{Accepted . Received ; in original form }

\pagerange{\pageref{firstpage}--\pageref{lastpage}} \pubyear{}

\maketitle

\label{firstpage}

\begin{abstract}
We study the global structure of optically thin, advection dominated, magnetized
accretion flow around black holes. We consider the magnetic field to be
turbulent in nature and dominated by the toroidal component. With this, we obtain
the complete set of accretion solutions for dissipative flows where bremsstrahlung
process is regarded as the dominant cooling mechanism. We show that rotating 
magnetized accretion flow experiences virtual barrier around black hole due to
centrifugal repulsion that can trigger the discontinuous transition of the flow
variables in the form of shock waves. We examine the properties of the
shock waves and find that the dynamics of the post-shock corona (PSC) is controlled
by the flow parameters, namely viscosity, cooling rate and strength of the
magnetic field, respectively. We separate the effective region of the parameter space for
standing shock and observe that shock can form for wide range of flow parameters.
We obtain the critical viscosity parameter that allows global accretion
solutions including shocks.
We estimate the energy dissipation at the PSC from where a part of the accreting matter
can deflect as outflows and jets. We compare the maximum energy that could be extracted
from the PSC and the observed radio luminosity values for several super-massive
black hole sources and the observational implications of our present analysis are discussed.
\end{abstract}

\begin{keywords}
accretion, accretion discs - black hole physics - hydrodynamics - shock waves
\end{keywords}

\section{Introduction}

In the quest of the accretion process around black holes, viscosity plays an important
role in a differentially rotating flow which is likely to be threaded by the magnetic
fields as well. Unfortunately, the source of the viscosity in an accretion disc is not
yet known conclusively. Meanwhile, \citet{Balbus-Hawley91,Balbus-Hawley98} showed that
viscosity  seems to arise in an accretion disc as a
consequence of the magneto-rotational instability (MRI). In this view, 
the Maxwell stress is generated by MRI that efficiently transports the 
angular momentum of the disc and at the same time the dissipation of the magnetic energy
is being utilized in disc heating \citep{Hirose-etal06}.
Based on the above consideration, 
several attempts were made to study the self-consistent global accretion solutions
around black holes \citep{Akizuki-Fukue06,Machida-etal06,Begelman-Pringle07,
Bu-etal09,Oda-etal07,Oda-etal10,Oda-etal12,Samadi-etal14}.
In these approaches, the description of the 
magnetic fields are considered to be toroidal in nature because the motion of the
accreting material inside the disc is primarily governed by the differential rotation
and therefore, the description of the magnetic field is expected to be dominated by 
the toroidal component of the magnetic fields. The work of \citet{Oda-etal10}
demonstrates the implication of the magnetically supported disc where it was shown
that the model has the potential to describe the bright/hard state observed during
the bright/slow transition of galactic black hole candidates. Observational signature
of such state transition was reported by \citet{Gierlinski-Newton06}.

In the conventional theory of the advective accretion disc around black holes,
sub-sonic inflowing matter starts its journey towards black hole from the outer
edge of the disc at large distance and in order to satisfy the inner boundary
conditions, flow must change its sonic state to become super-sonic before
crossing the horizon. In the vicinity of the black hole, rotating flow experiences
centrifugal barrier against gravity that eventually triggers the discontinuous
transition of the flow variables in the form of shock waves when possible.
According to the second law of thermodynamics, accretion solutions containing 
shock waves are preferred as they possess high entropy content
\citep{Becker-Kazanas01}.
Presence of shock waves in an accretion disc around black hole has been confirmed
both theoretically
\citep{Fukue87,Chakrabarti89,Chakrabarti96,Das-etal01a,Chakrabarti-Das04,Lu-etal99} as well
as numerically \citep{Molteni-etal94,Molteni-etal96,Das-etal14,Okuda14,Okuda-Das15}.
Due to shock compression, the post-shock matter, equivalently post-shock corona
(PSC), becomes hot and dense compared to the pre-shock matter and eventually PSC
behaves like an effective boundary layer of the black hole. Since PSC is composed
with the swarm of hot electrons, soft-photons from the cold pre-shock matter are
inverse Comptonized after intercepted at the PSC and produces the spectral 
features of the black holes \citep{Chakrabarti-Titarchuk95}. In addition,
PSC deflects a part of the accreting matter to produce bipolar jets and outflows
due to the excess thermal gradient force present across the shock
\citep{Chakrabarti99,Das-etal01b,Chattopadhyay-Das07,Das-Chattopadhyay08,Aktar-etal15}.
When PSC modulates,
quasi-periodic oscillation (QPO) of hard radiations in the spectral states is observed
\citep{Chakrabarti-Manickam00,Nandi-etal01a,Nandi-etal01b,Nandi-etal12}
along with the variable outflow rates \citep{Das-etal01b}.

Although the shocked accretion solutions seems to have potential to describe
the spectral and timing properties as well as outflow rates, 
no efforts are given to examine the properties of the magnetically supported
accretion flow that harbors shock waves. Being motivated with this, in the present
paper, we model the optically thin magnetized accretion flow around a
Schwarzschild black hole. The characteristic of the magnetic pressure is assumed to be
same as the gas pressure and their combined effects support the vertical structure
of the disc against gravity. Following the conventional $\alpha$-viscosity
prescription of \citet{Shakura-Sunyaev73}, it is evident that the angular momentum
transport in the disc
equatorial plane would also be enhanced as the magnetic pressure contributes
to the total pressure. Towards this,
we consider a set of steady state hydrodynamical equations describing the dissipative
accretion flow in a disc. For simplicity, the space-time geometry around a
Schwarzschild black hole is approximated by adopting the pseudo-Newtonian potential
\citep{Paczynski-Wiita80}.
We further consider that the heating of the flow is governed by the magnetic energy
dissipation process and the cooling of the flow is dominated by the 
Comptonization of the bremsstrahlung
radiation, respectively. With this, we self-consistently calculate the
global accretion solution including shock waves and investigate the shock properties
in terms of the flow parameters. We find that shocked accretion
solution exists for a wide range of flow parameters.
We also computed the critical value of viscosity parameter
$\alpha^{\rm cri}_{B}$ for which standing shock forms in a magnetized flow.
Indeed, $\alpha^{\rm cri}_{B}$ greatly depends on the inflow parameters.
Note that $\alpha^{\rm cri}_{B}$ tends to $\alpha_{\Pi}^{\rm cri}(\sim 0.3)$ as
estimated by \citet{Chakrabarti-Das04} for gas pressure dominated flow.
This is quite obvious because in the adopted
viscosity prescription, magnetic pressure contributes to the total pressure and
hence, a lower value of $\alpha_{B}$ is sufficient to transport the required angular
momentum for shock transition.
This essentially establishes the fact that the shocks under consideration are
centrifugally driven. Further more, we consider the shock to be dissipative in nature
and compute the maximum available energy dissipated at the shock.
Employing this result, we then calculate the loss of kinetic power from
the disc ($L^{\rm max}_{\rm shock}$) which could be utilized to power the jets as
they are likely to be launched from the PSC 
\citep{Chakrabarti99,Das-etal01b,Das-Chattopadhyay08,Aktar-etal15}. The above analysis
apparently provides an estimate which we compare with the jet kinetic power
available from observation for six sources and close agreements are
seen. 

The plan of the paper is as follows. In the next Section, we describe the
assumptions and governing equations for our model. In Section 3,
we present the global accretion solutions with and without shock, shock properties,
and shock parameter space. In Section 4, we apply our formalism to calculate the
shock luminosity considering several astrophysical sources. In section 5, we
present concluding remarks.

\section{Governing Equations}

We begin with the consideration that the magnetic fields inside the accretion disc
are turbulent in nature and the azimuthal component of the magnetic fields dominates
over other component. Numerical study of global MHD accretion flow around black holes
in the quasi-steady state supports the above findings \citep{Machida-etal06,Hirose-etal06}.
Based on the simulation work, the magnetic fields are considered as a combination of mean
fields and the fluctuating fields. The mean fields are denoted as ${\vec B}=(0,<B_{\phi}>,0)$
where, $<>$ indicates the azimuthal average and the fluctuating fields are represented by
$\delta {\vec B}= (\delta {B}_{r}, \delta {B}_{\phi},\delta {B}_{z})$. When the fluctuating
fields are averaged azimuthally, we assume that they eventually disappear. Therefore, the
azimuthal component of magnetic fields dominates over the other components as they are
negligible, $\mid <B_{\phi}> + \delta B_{\phi}\mid \gg \mid \delta B_{r} \mid ~{\rm and}
\mid \delta B_{z}\mid$. This essentially yields the azimuthally averaged magnetic field as
$<\vec B>=<B_{\phi}> \hat{\phi}$ \citep{Oda-etal07}.

In this work, we use geometric units as $2G=M_{BH}=c=1$, where $G$, $M_{BH}$ and $c$
are the gravitational constant, mass of the black hole and the speed of light, 
respectively. In this unit system, length, time and velocity are expressed in
unit of $r_g=2GM_{BH}/c^2$, $2GM_{BH}/c^3$ and $c$, respectively. Here,
we assume that the matter accretes through the equatorial plane of a Schwarzschild black
hole. We use cylindrical polar coordinates ($x,\phi,z$) with the black hole at the
origin and the disc lies in the $z = 0$ plane. We adopt the pseudo-Newtonian potential
\citep{Paczynski-Wiita80} to describe the space-time geometry around the black hole
and is given by, 

$$
\Phi = -\frac{1}{2(x -1)},
\eqno(1)
$$
where, $x$ is the non-dimensional radial distance.

The gas pressure 
inside the disc is obtained as $p_{gas} = R \rho T/\mu$, where, $R$ is the
gas constant, $\rho$ is the density, $T$ is the temperature and $\mu$ 
is the mean molecular weight assumed to be $0.5$ for fully ionized hydrogen.
The magnetic pressure is given by, $p_{mag} = <B_{\phi}^2>/8\pi$, where, $<B_{\phi}^2>$ 
is the azimuthal average of the square of the toroidal component of the magnetic field. 
We denote the total pressure in the disc by $P=p_{gas}+p_{mag}$. We define plasma $\beta$
as the ratio of gas pressure ($p_{gas}$) to the magnetic pressure ($p_{mag}$) inside
the disc which yields $P = p_{gas} (1 + 1/\beta)$. 
The adiabatic sound speed is defined as $a=\sqrt {\gamma P/\rho}$, where $\gamma$ is 
the adiabatic index assumed to be constant throughout the flow. We adopt the canonical value
of $\gamma=1.5$ in the subsequent analysis. We consider the disc to be axisymmetric,
steady and thin. Following this, we compute the half thickness of the disc
($h$) considering the flow is in hydrostatic equilibrium in the transverse direction
and is given by, 
$$
h = \sqrt{\frac{2}{\gamma}} a x^{1/2} (x - 1).
\eqno(2)
$$

With this, we have the following governing equations that describes the accreting
matter in the steady state as:

\noindent (a) Radial momentum equation:

$$
{v\frac{dv}{dx} + \frac{1}{\rho}\frac{dP}{dx} - \frac{\lambda^2(x)}{x^3} 
+ \frac{1}{2(x-1)^2} + \frac{\left<B_{\phi} ^2\right>}{4\pi x \rho} = 0}.
\eqno(3)
$$
where, $v$ is the radial velocity and $\lambda(x)$ is the specific angular momentum
at radial coordinate $x$. The last term on the left hand side represents the
magnetic tension force. 

\noindent (b) Mass Conservation:

$$
\dot{M}=2\pi x\Sigma v,
\eqno(4)
$$
where, $\dot{M}$ is the rate at which the black hole is continuously 
accreting matter and remain constant throughout the flow. $\Sigma$
represent the vertically integrated density of flow
\citep{Matsumoto-etal84}.

\noindent (c) Azimuthal momentum equation:

$$
v\frac{d\lambda(x)}{dx}+\frac{1}{\Sigma x}\frac{d}{dx}(x^2T_{x\phi}) = 0,
\eqno(5)
$$
where, we consider that the vertically integrated total stress is dominated by the $x\phi$
component of the Maxwell stress $T_{x\phi}$. Following \citet{Machida-etal06},
we estimate $T_{x\phi}$ for an advective flow with significant radial velocity
\citep{Chakrabarti-Das04} as
$$
T_{x\phi} = \frac{<B_{x}B_{\phi}>}{4\pi}h = -\alpha_{B}(W + \Sigma v^2),
\eqno(6)
$$
where, $\alpha_B$ is the proportionality constant and $W$ is the vertically 
integrated pressure \citep{Matsumoto-etal84}. In the
present study, we treat $\alpha_B$ as
a parameter based on the seminal work of \citet{Shakura-Sunyaev73}.
For a Keplerian flow where the radial velocity is unimportant, Eq. (6) subsequently
reduces to the original prescription of `$\alpha$-model' \citep{Shakura-Sunyaev73}.

\noindent (d) The entropy generation equation:

$$
\Sigma v T \frac {ds}{dx}=\frac{hv}{\gamma-1}
\left(\frac{dP}{dx} -\frac{\gamma P}{\rho}\frac{d\rho}{dx}\right)=Q^- - Q^+,
\eqno(7)
$$
where, we consider $\beta>1$ inside the flow.
Subsequently, we assume $\beta/(\beta+1)\sim 1$ and neglect term with
$1/(\beta+1)^2$ for a modest value of $\beta$.
Here, $s$ and $T$ represent the specific entropy and the local temperature of
the flow, respectively. 
In the right hand side, $Q^+$ and $Q^-$ denote the
vertically integrated heating and cooling rates. The
flow is heated due to
the thermalization of magnetic energy through the magnetic reconnection
mechanism \citep{Hirose-etal06,Machida-etal06} and therefore, expressed as
$$
Q^{+} = \frac{<B_{x}B_{\phi}>}{4\pi} x h \frac{d\Omega}{dx} = 
-\alpha _{B}(W + \Sigma v^2) x \frac{d\Omega}{dx},
\eqno(8)
$$
where, $\Omega$ denotes the angular velocity of the flow.

The cooling of the flow could be due to the various physical processes, namely
bremsstrahlung, synchrotron, Comptonization etc. For simplicity, in this work, we
assume the Comptonization of the bremsstrahlung radiation where the intensity
of the bremsstrahlung photons are enhanced by a factor $\xi$. Evidently, 
$1<\xi<{\sim {\rm few} \times 100}$, depending on the availability of soft
photons \citep{Das-Chakrabarti04,Chakrabarti-Titarchuk95}. In a way, $\xi$ is
treated as dimensionless parameter in the form of cooling efficiency factor
to represent net cooling. When $\xi = 0$, flow becomes heating dominated as
it cools inefficiently. In this view, the cooling rate of the flow is given by 
\citep{Shapiro-Teukolsky83}, 
$$
Q^-=\xi \times \frac{C\rho h}{vx^{3/2}(x-1)} \left[\frac{\beta}{1+\beta}\right]^{\frac{1}{2}},
\eqno(9)
$$
with,
$$
C=1.974 \times 10^{-10} \left(\frac{m_e}{m_p}\right)^{\frac{1}{4}}
\left[\frac{\mu m_p}{2k_B}\right]^{\frac{1}{2}}\frac{\dot m}{4\pi I_n m_p^2}
\frac{1}{2GcM_{\odot}},
$$ 
where, $m_p$ is the mass of the ion, $m_e$ is the mass of the electron, $k_B$ is the
Boltzmann constant, $I_n = (2^n n!)^2/(2n + 1)!$ and $n = 1/(\gamma - 1)$. In this analysis,
we ignore any coupling between the ions and electrons and estimate the electron
temperature using the relation $T_e = \sqrt{m_e/m_p}T_p$ \citep{Chattopadhyay-Chakrabarti02}.
Further more, here ${\dot m}$ represents the accretion rate measured in units of Eddington
rate and we consider ${\dot m} =0.05$ all throughout the paper until otherwise stated.

\noindent (e) Radial advection of the toroidal magnetic flux:

In order to describe the advection rate of the toroidal magnetic flux we consider the 
induction equation which is given by,

$$
\frac {\partial <B_{\phi}>\hat{\phi}}{\partial t} = {\bf \nabla} \times
\left({\vec{v}} \times <B_{\phi}>\hat{\phi} -{\frac{4\pi}{c}}\eta {\vec{j}}\right),
\eqno(10)
$$
where, $\vec {v}$ is the velocity and
${\vec{j}} = c\left({\nabla} \times <B_{\phi}>\hat{\phi}\right)/4\pi$
is the current density. Here, Eq. (10) is azimuthally averaged
and the dynamo and magnetic-diffusion terms are neglected. In
the steady state, the resulting
equation is then vertically averaged considering the fact that the averaged toroidal
magnetic fields vanish at the surface of the disc. This yields the advection rate
of the toroidal magnetic flux as \citep{Oda-etal07},
$$
\dot{\Phi} = - \sqrt{4\pi}v h {B}_{0} (x),
\eqno(11)
$$
where,
\begin{eqnarray*}
{B}_{0} (x) && = \langle {B}_{\phi} \rangle \left(x; z = 0\right)  \nonumber \\
&& = 2^{5/4}{\pi}^{1/4}(R T/\mu)^{1/2}{\Sigma}^{1/2}h^{-1/2}{\beta}^{-1/2}
\end{eqnarray*}
is the azimuthally averaged toroidal magnetic field lies in the disc equatorial plane.
According to Eq. (10), $\dot{\Phi}$ is expected to vary with radial coordinate of the
accretion disc due to the presence of the dynamo term and the magnetic diffusion term.
Meanwhile, \citet{Machida-etal06} numerically showed out that $\dot{\Phi} \propto x^{-1}$,
when the disc is in quasi steady state. Following this result, we adopt a parametric
relation between $\dot{\Phi}$ and $x$ which is given by \citep{Oda-etal07}

$$
\dot{\Phi}\left(x; \zeta, \dot{M}\right) \equiv \dot{\Phi}_{\rm edge}(\dot{M})
\left(\frac{x}{x_{\rm edge}} \right)^{-\zeta},
\eqno(12)
$$
where, $\dot{\Phi}_{edge}$ denotes the advection rate of the toroidal magnetic flux
at the outer edge of the disc ($x_{edge}$). Note that the conservation of the magnetic
flux is restored when $\zeta = 0$. However, for $\zeta > 0$, the magnetic flux
increases as the
accreting matter proceeds towards the black hole horizon. In this work, we consider $\zeta$
to remain constant all throughout and adopt $\zeta =1$ for representation, until
otherwise stated.

\subsection{Sonic Point Analysis}

In order to study the dynamical structure of the accretion flow, one needs to obtain
the global accretion solution where infalling matter from the outer edge of the disc
can smoothly accrete inwards before entering in to the black hole. In addition, it
is necessary for the accretion solution to become transonic in nature in order to 
satisfy the inner boundary conditions imposed by the black hole event horizon. Based
on the above insight, we visualize the general nature of the sonic points by solving 
Eqs. (3-7) and Eqs. (11-12) simultaneously \citep{Das07} which is expressed as,

$$
\frac {dv}{dx}=\frac{N}{D},
\eqno(13)
$$
where, the numerator ($N$) is given by,

$$
N =\frac {C}{v x^{3/2}(x-1)}\frac{\beta^{1/2}}{(1 + \beta)^{1/2}}
+\frac {2\alpha^2_B I_n  (a^2g+\gamma v^2)^2}{\gamma^2 x v}
$$
$$
+\frac {2\alpha^2_B g I_n a^2(5x-3)(a^2g+\gamma v^2)}{\gamma^2 v x(x-1)}
$$
$$
-\left[ \frac {\lambda^2}{x^3}-\frac {1}{2(x-1)^2}\right]
\left[\frac {(\gamma+1)v}{(\gamma-1)}
-\frac{4\alpha^2_B g I_n (a^2g+\gamma v^2)}{\gamma v} \right]
$$
$$
- \frac {v a^2(5x-3)}{x(\gamma-1)(x-1)} -\frac {4\lambda \alpha_B I_n
(a^2g+\gamma v^2)}{\gamma x^2}
$$
$$
 -\frac {8 \alpha_B^2 I_n a^2 g(a^2g+\gamma v^2)}{\gamma^2 v (1 + \beta) x} 
+\frac {2(\gamma + 1)a^2 v}{\gamma(\gamma - 1)(1 + \beta) x}
\eqno(13a)
$$

and the denominator $D$ is,
$$
D = \frac {2a^2}{(\gamma-1)}-\frac {(\gamma+1)v^2}
{(\gamma-1)}
$$
$$
+\frac{2\alpha^2_B I_n (a^2g+\gamma v^2)}{\gamma}
\left[ (2g-1)-\frac {a^2g}{\gamma v^2}\right].
\eqno(13b)
$$
Here, we write $g=I_{n+1}/I_{n}$.

The gradient of sound speed is calculated as,

$$
\frac{da}{dx}=\left( \frac{a}{v} - \frac{\gamma v}{a} \right)
\frac{dv}{dx} + \frac{\gamma}{a}\left[ \frac {\lambda^2}{x^3}-\frac {1}{2(x-1)^2}\right]
$$
$$
+\frac{(5x-3)a}{2x(x-1)} - \frac{2a}{(1+\beta)x}
\eqno(14)
$$

The gradient of angular momentum is obtained as,

$$
\frac{d\lambda}{dx}=
-\frac{\alpha_{B} x (a^2g- \gamma v^2)}{\gamma v^2}\frac{dv}{dx}
+\frac{2 \alpha_{B} axg }{\gamma v}\frac{da}{dx}
$$
$$
+\frac{\alpha_{B}(a^2g+\gamma v^2)}{\gamma v}
\eqno(15)
$$

The gradient of plasma $\beta$ is given by:

$$
\frac{d\beta}{dx}= \frac{(1+\beta)}{v}\frac{dv}{dx}
+\frac{3 (1+\beta)}{a}\frac{da}{dx}+\frac{1+\beta}{x-1}
$$
$$
+\frac{(1+\beta)(4\zeta-1)}{2x}
\eqno(16)
$$

Matter starts accreting towards the black hole from the outer edge of the disc with
almost negligible velocity and subsequently crosses the black hole horizon with velocity
equal to the speed of light. This suggests that the accretion flow trajectory must be
smooth along the streamline and therefore, the radial velocity gradient would be
necessarily real and finite always. However, Eq. (13b) indicates that there may be
some points between the outer edge of the disc and the horizon, where the denominator
($D$) vanishes. To maintain the flow to be smooth everywhere along the streamline, 
the point where $D$ tends to zero, $N$ must also vanish there. The point where both
$N$ and $D$ vanish simultaneously is a special point and called as sonic point
($x_c$). Thus, we have $N=D=0$ at the sonic point. Setting $D=0$, we obtain the
expression of the Mach number ($M=v/a$) at the sonic point which is calculated as,

$$
M_{c} =\sqrt {\frac{-m_b - \sqrt{m^2_b-4m_a m_c}}{2m_a}},
\eqno(17)
$$
where,
$$ 
m_a=2\alpha^2_{B} I_n \gamma(\gamma-1)(2g-1) - \gamma(\gamma+1)
$$
$$
m_b=2\gamma + 4\alpha^2_{B} I_n g (g-1)(\gamma-1)
$$ 
$$
m_c=-(2\alpha^2_{B} I_n g^2 (\gamma-1))/\gamma
$$

Setting $N=0$, we obtain the algebraic equation of the sound speed at the sonic
point and is given by,

$$
{\mathcal A}_{c}a^4(x_{c}) + {\mathcal B}_{c}a^3(x_{c})
+ {\mathcal C}_{c}a^2(x_{c}) +{\mathcal D}_{c}= 0 ,
\eqno(18)
$$
where,
$$
{\mathcal A}_{c} = \frac {2\alpha^2_B I_n (g+\gamma M_{c}^2)^2}
{\gamma^2 x_{c}}+\frac {2\alpha^2_B I_n g (5x_{c}-3)(g+\gamma M_{c}^2)}
{\gamma^2 x_{c}(x_{c}-1)} 
$$
$$
-\frac{M_{c}^2(5x_{c}-3)}{x_{c}(\gamma-1)(x_{c}-1)} -\frac {8\alpha^2_B I_n g(g+\gamma M_{c}^2)}
{\gamma^2 (1+\beta_{c})x_{c}}
$$
$$
 + \frac{2(\gamma+1)M_{c}^2}{\gamma(\gamma-1)(1+\beta_{c})x_{c}},
$$

$$
{\mathcal B}_{c} = -\frac {4\lambda_{c} \alpha_B I_n M_{c} (g+\gamma M_{c}^2)}{\gamma x_{c}^2},
$$

$$
{\mathcal C}_{c} = -\left[ \frac {\lambda_{c}^2}{x_{c}^3}-\frac {1}{2(x_{c}-1)^2}\right]
$$
$$
\times \left[\frac {(\gamma+1) M_{c}^2}{(\gamma-1)}-\frac{4\alpha^2_B g I_n (g+\gamma M_{c}^2)}{\gamma} \right],
$$
and

$$
{\mathcal D}_{c}=\frac{C}{x_{c}^{3/2}(x_{c}-1)}\sqrt{\frac{\beta_{c}}{1+\beta_{c}}}
$$
Here, the subscript `c' denotes the flow variables at the sonic point.

We solve Eq. (18) to calculate the sound speed at the sonic point knowing the input
parameters of the flow and subsequently, we obtain the radial velocity at the sonic
point from Eq. (17). Following this, it is straight forward to study the properties
of the sonic points and its classification through the extensive investigation of Eq. (13).
At the sonic point, $dv/dx$ generally owns two distinct values corresponding to
accretion and wind solutions. When both the derivatives are real and of opposite
sign, the sonic point is considered to be a matter of special interest as the global
transonic solutions only pass through it and such a point is called as saddle type
sonic point \citep{Chakrabarti-Das04}. In this work, our main focus is to examine
the dynamical structure of accretion flow and its various properties and therefore,
the wind solutions are left aside.

\section{Global Accretion Solution}

\begin{figure}
\begin{center}
\includegraphics[width=0.5\textwidth]{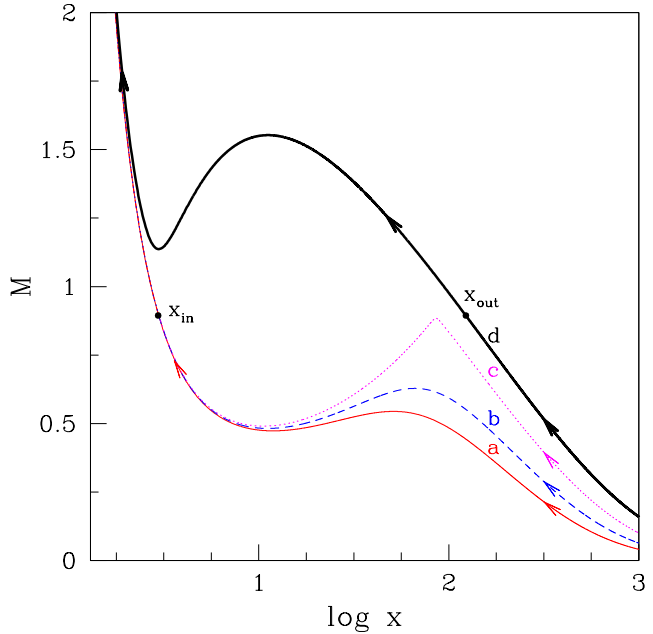}
\end{center}
\caption{Radial dependence of Mach number ($M=v/a$) of the accreting matter for
different values of angular momentum ($\lambda_{\rm edge}$) at the outer edge
$x_{\rm edge} = 1000$ where $\beta_{\rm edge} = 1400$, $\alpha_B=0.01$ and 
$\xi = 10$.
Thin solid and dashed curves represent the results for $\lambda_{\rm edge}=5.9459$
and $4.1443$, respectively. For the same set of outer edge parameters, 
the minimum angular momentum that provides the accretion solution passing through
the inner sonic is identified as $\lambda^{\rm min}_{\rm edge}=3.0619$ (dotted curve).
When $\lambda_{\rm edge} < \lambda^{\rm min}_{\rm edge}$, accretion solutions pass through
the outer sonic point only (thick solid curve) where $\lambda_{\rm edge}=2.4021$.
In the figure, the locations of the inner sonic point ($x_{\rm in}$) and outer sonic
point ($x_{\rm out}$) are marked and arrows indicate the direction of the flow
towards the black hole. See text for details.
}
\end{figure}

In order to obtain a global accretion solution, we solve Eqs. (13-16) simultaneously
knowing the boundary values of angular momentum ($\lambda$), plasma $\beta$, cooling efficiency 
factor ($\xi$) and $\alpha_{B}$ at a given radial distance $(x)$. Since the black
hole solutions are necessarily transonic, flow must pass through the sonic point and 
therefore, it is convenient to supply the boundary values of the flow at the sonic point.
With this, we integrate Eqs. (13-16) from the sonic point once inward up to the black
hole horizon and then outward up to a large distance (equivalently `disc outer edge')
and finally join them to obtain a complete global transonic accretion solution. Depending
on the input parameters, flow may possess single or multiple sonic points
\citep{Das-etal01a}. When the 
sonic points form close to the horizon, they are called as inner sonic points ($x_{\rm in}$)
and when they form far away from the horizon, they are called as outer sonic
points ($x_{\rm out}$), respectively.

\subsection{Shock Free Global Accretion Solution}

In Fig. 1, we present the examples of accretion solutions where the variation
of Mach number $(M=v/a)$ is plotted as function of logarithmic radial distance ($x$).
The solid curve
marked `a' represents a global accretion solution passing through the inner sonic point
$x_{\rm in}=2.9740$ with angular momentum $\lambda_{\rm in}=1.4850$,
$\beta_{\rm in}=27.778$, $\alpha_{B}=0.01$ and $\xi=10$, respectively and connects
the BH horizon with the outer edge of the disc $x_{\rm edge}$ where we note the values
of the flow variables $\lambda_{\rm edge} = 5.9459$, $\beta_{\rm edge} = 1400$,
$v_{\rm edge}=0.00132$, $a_{\rm edge}=0.03205$ at $x_{\rm edge} = 1000$. 
Alternatively, one can obtain the same solution
when the integration is carried out towards the black hole starting from the outer
edge of the disc ($x_{\rm edge}$) with the noted boundary values.
Hence, the above result essentially represents the solution of an accretion flow that starts
its journey from $x_{\rm edge} = 1000$ and crosses the inner sonic point at
$x_{\rm in}=2.9740$ before entering into the black hole. The arrow indicates the direction
of the flow.  Now, we decrease $\lambda_{\rm edge} = 4.1443$ 
keeping all the other values of the flow variables same at $x_{\rm edge} = 1000$ and
obtain the global transonic solution by suitably adjusting the values of
$v_{\rm edge}=0.00189$ and $a_{\rm edge}=0.02962$. The solution is marked as `b'. Here, 
the values of $v_{\rm edge}$ and $a_{\rm edge}$ is required additionally to start the
integration as the sonic point is not known apriori. 
Following this approach, we identify the minimum value of angular momentum at the outer edge 
$\lambda^{\rm min}_{\rm edge} = 3.0619$, below this value accretion solution fails to pass
through the inner sonic point. Accretion solution corresponding to the minimum
$\lambda^{\rm min}_{\rm edge}$ is indicated by the dotted curve and marked as `c'. 
The results namely `a-c' represent solutions similar to the solution of advection
dominated accretion flow (ADAF) around black holes \citep{Narayan-etal97,Oda-etal07},
although another important
class of solutions still remains unexplored which we present in this work.
As $\lambda^{\rm min}_{\rm edge}$ is decreased further, such as 2.4021,
accretion solution changes its character
and passes through the outer sonic point ($x_{\rm out}=122.9$) instead of inner sonic point
($x_{\rm in}$) with angular momentum ($\lambda_{\rm out}=1.5631$),
$\beta_{\rm out}=431.8$ which is indicated by the thick solid line marked as `d'.
In the frame work of magnetically supported accretion disc, accretion solution
passing through the outer sonic points was not studied so far. Solutions particularly
of this kind are potentially interesting as they may possess centrifugally supported
shock waves. The presence of shock wave in an accretion flow
has profound implications as it satisfactorily delineates the spectral and temporal
behaviour of numerous black hole sources \citep{Chakrabarti89,Chakrabarti90,
Chakrabarti96,Molteni-etal94,Molteni-etal96,Becker-Kazanas01,Lu-etal99,Das-etal01a,
Le-Becker04,Gu-Lu04,Le-Becker05,Chakrabarti-Das04,Becker-etal08,Nagakura-Yamada09,
Nandi-etal12,Das-etal09,Das-etal14,Okuda14,Iyer-etal15,Okuda-Das15,Aktar-etal15,
Sukova-Janiuk15}.
Thus, in this work we intend to study the properties of magnetically supported
accretion solutions that possesses shock waves.

\subsection{Shock Induced Global Accretion Solution}

\begin{figure}
\begin{center}
\includegraphics[width=0.5\textwidth]{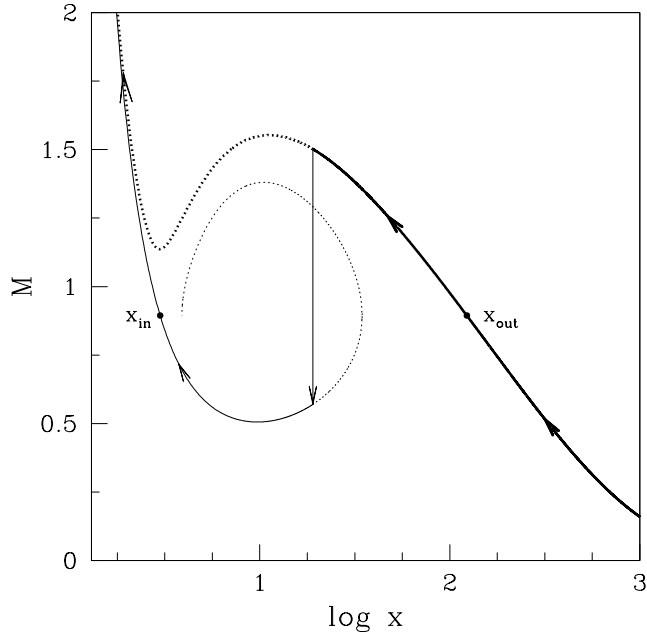}
\end{center}
\caption{\label{lab:fig1} A complete global accretion solution containing shock
($x_s = 19.03$) is depicted along with outer ($x_{\rm out}$) and inner ($x_{\rm in}$)
sonic points. Inflow parameters at the outer edge are same as case (d) of Fig. 1. See text
for details.
}
\end{figure}

In Fig. 2, we present a global accretion solution that contains shock wave where the 
flow crosses the sonic region multiple times. 
Here, we consider inflowing matter that starts accreting towards the black
hole sub-sonically with the boundary values at the outer edge same as the
case `d' of Fig. 1 and becomes supersonic after crossing the outer sonic point at
$x_{\rm out}=122.9$. As the rotating matter proceeds further, it experiences virtual
barrier due to centrifugal repulsion and starts piling up there. The process continues
and at some point, the flow eventually encounters discontinuous transition
of flow variables in the form of shock when shock conditions are satisfied. 
This is because the shock solutions are thermodynamically preferred
as the post-shock matter possesses high entropy content \citep{Becker-Kazanas01}.
Following  \citet{Landau-Lifshitz59},
the conditions for shock transition in a vertically averaged flow are considered as
the conservation of (a) mass flux (${\dot M}_{-}={\dot M_{+}}$) (b) the momentum flux
($W_{-}+\Sigma_{-} v^2_{-}=W_{+}+\Sigma_{+} v^2_{+}$) (c) the energy flux, obtained
integrating Eq. (3) (${\cal E_{-}}={\cal E_{+}}$) and (d) the magnetic flux 
($\dot {\Phi}_{-}=\dot {\Phi}_{+}$) across the shock. Here, the quantities having
subscripts `-' and `+' are referred to the values before and after the shock.
While doing so, we assume the shock to be thin and non-dissipative. 
In the post-shock region, flow momentarily slows down as it becomes subsonic immediately
after the shock transition and the pre-shock kinetic energy is then converted in to the thermal
energy. Therefore, the post-shock matter essentially becomes hot and dense.
Due to gravitational attraction, subsonic post-shock matter continues to accrete towards the BH
and gradually picks up its radial velocity and 
subsequently crosses the inner sonic point smoothly in order to satisfy the supersonic
inner boundary condition before jumping in to the black hole. In the figure,
we depict the variation of Mach number with the logarithmic radial distance. Thick curve
denotes the accretion solution passing through the outer sonic point which in principle 
can enter in to the black hole directly. Interestingly, on the way towards the black hole, as
the shock conditions are satisfied, flow makes discontinuous jump from the supersonic
branch to the subsonic branch avoiding thick dotted part of the solution.
In the figure, the joining of the supersonic pre-shock flow with the subsonic post-shock flow is
indicated by the vertical arrow and 
the thin solid line denotes the inner part of the solution representing the
post-shock flow. Here, $x_{\rm in}$ and $x_{\rm out}$ are the inner and outer sonic points,
respectively. Arrows indicate the overall direction of the flow motion during accretion
towards black hole.

\begin{figure}
\begin{center}
\includegraphics[width=0.45\textwidth]{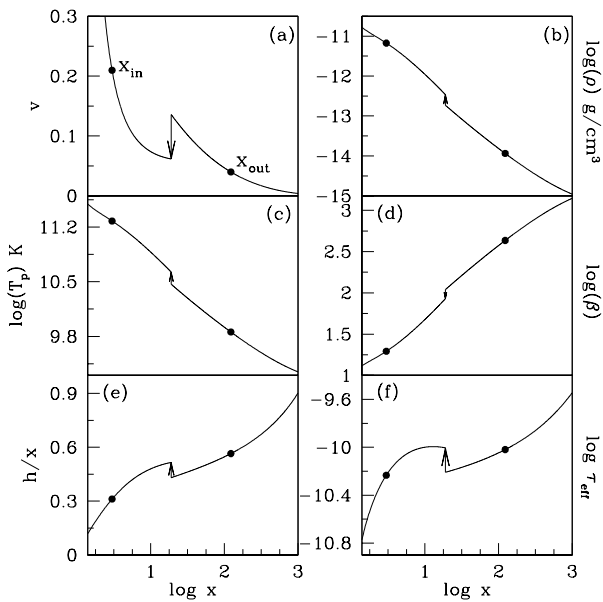}
\end{center}
\caption{Variation of (a) radial velocity, (b) density in g/cm$^3$,
(c) Temperature (d) ratio of gas pressure to magnetic pressure, (e)
disc scale height ($h/x$) and (f) effective optical depth as function
of radial coordinate around a Schwarzschild black hole. All the flow variables correspond
to the solution depicted in Fig. 2. Filled circles represent the sonic points where the
closer one is the inner sonic point and the furthest one is the outer sonic point.
Vertical arrows indicate the location of the shock location $x_s = 19.03$. See text for
details.
}
\end{figure}

In Fig. 3, we study the structure of a vertically averaged accretion disc corresponding 
to the solution depicted in Fig. 2. Here, each panel shows the variation of flow variables
as function of logarithmic radial distance. In Fig. 3a, we demonstrate the radial
velocity ($v$) variation of the accreting flow where the shock transition is observed at 
($x_s=19.03$) indicated by the vertical arrow. In Fig. 3b, we show the density profile of
the flow where the catastrophic jump of density at the shock location is observed. This
happens mainly due
to the reduction of radial velocity in the post-shock flow where the conservation of mass
accretion is preserved across the shock. 
The formation of shock causes the compression of the post-shock flow that along with the
enhancement of density effectively increases the temperature of the flow at the inner part of the
disc which we represent in Fig. 3c. We display the variation of plasma $\beta$ in Fig. 3d
where a noticeable reduction of $\beta$ is seen at the shock location. In Fig. 3e, we present
the dependence of the vertical scale-height ($h/x$) on the radial coordinate. Here, we observe
that the half thickness of the disc always remain smaller than the local radial coordinate
all the way from the outer edge of the disc to the horizon even in presence of shock wave.
We estimate the effective optical depth as
$\tau_{\rm eff} = \sqrt{\tau_{\rm es}\tau_{\rm br}}$ where, $\tau_{\rm es}$ denotes the
scattering optical depth given by $\tau_{\rm es} = \kappa_{\rm es} \rho h$ and the
electron scattering opacity is given by $\kappa_{\rm es} = 0.38~{\rm cm}^2 {\rm g}^{-1}$.
Here, $\tau_{\rm br}$ represents the absorption effect appears due to 
thermal processes and is given by 
$\tau_{\rm br} =\left( h q_{\rm br}/4 \sigma T_{e}^4\right)\left(2GM_{BH}/c^2\right)$
where, $q_{\rm br}$ is the
bremsstrahlung emissivity \citep{Shapiro-Teukolsky83} and $\sigma$ is the Stefan-Boltzmann
constant. For the purpose of representation, here we consider $M_{BH} = 10^6 M_{\odot}$.
We find that the post-shock flow remain optically thin ($\tau < 1$) although
the density profile is steeper there. This intuitively suggests that the possibility 
of escaping hard radiations from the PSC would be quite significant.

\subsection{Shock Dynamics and Properties}

\begin{figure}
\begin{center}
\includegraphics[width=0.45\textwidth]{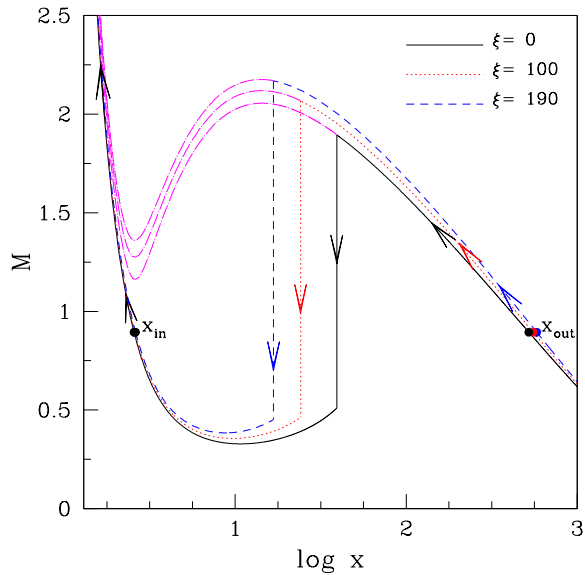}
\end{center}
\caption{Plot of Mach number with logarithmic radial distance for different values
of cooling factor ($\xi$). Accreting flows are injected from $x_{\rm edge} = 1000$
with $\lambda_{\rm edge} = 1.88$, ${\cal E}_{\rm edge}=1.9133 \times 10^{-4}$,
$\beta_{\rm edge} = 500$, and $\alpha_B = 0.01$. Solution obtained for cooling free case
($\xi=0$) is denoted by the solid curve whereas dotted and dashed curves represent the solution
for $\xi = 100$ and $190$, respectively. The
corresponding shock locations are
indicated by the vertical arrow as $x_s = 39.27$ (solid), $24.20$ (dotted)
and $16.78$ (dashed). Sonic points are marked by the filled circles.
}
\end{figure}

Here, we examine the effect of cooling on the dynamical structure of the
accretion flow that contain shock waves. In order for that
we fix the outer edge of the disc at $x_{\rm edge} = 1000$ and inject matter
sub-sonically with local angular momentum $\lambda_{\rm edge} = 1.88$,
$\beta_{\rm edge} = 500$, ${\cal E}_{\rm edge}=1.9133 \times 10^{-4}$ and
$\alpha_B = 0.01$, respectively. First, we consider a
cooling free flow ($\xi=0$) that becomes supersonic after crossing the outer sonic point
($x_{\rm out}=521.22$) and continues its journey
towards the black hole. Meanwhile, stationary shock conditions are satisfied and accreting
matter encounters a shock transition depicted in Fig. 4 where Mach number ($M$) of the flow is
plotted as function of logarithmic radial coordinate. The solid vertical arrow indicates
the location of the standing shock at $x_s = 39.27$ for flows having no cooling. Next,
we introduce cooling considering the flow parameters at the outer edge same as in the cooling
free case. When cooling efficiency factor $\xi = 100$ is supplied, shock forms at $x_s = 24.20$
indicated by the dotted vertical arrow.
In reality, due to shock compression, the density and temperature in the post-shock flow
are enhanced compared to the pre-shock flow and therefore, cooling is very much effective there
that reduces the post-shock pressure significantly.
This causes the shock front to
move forward towards the horizon in order to maintain the pressure balance
on either sides of the shock. This clearly indicates that the dynamics of the shock in a way
are controlled by the resultant pressure across it. With the gradual
increase of the cooling factor $\xi$, shock front proceeds closer to the BH horizon.
Following this, we identify the extreme value of cooling factor $\xi = 190$ that 
provides the global accretion solution including shock waves at $x_s = 16.78$ for the
same outer boundary parameters as considered in cooling free case. The shock
location for $\xi = 190$ is represented by the dashed vertical line in the figure.
When $\xi$ is increased further, 
shocked accretion solution ceases to exist as the shock conditions are not satisfied there.
Note that we obtain the shock induced global accretion solution even for very high
cooling efficiency factor. This is possible because the effect of bremsstrahlung
cooling in an accretion flow is normally weak as pointed out by 
\citet{Chattopadhyay-Chakrabarti00,Das-Chakrabarti04}.

\begin{figure}
\begin{center}
\includegraphics[width=0.45\textwidth]{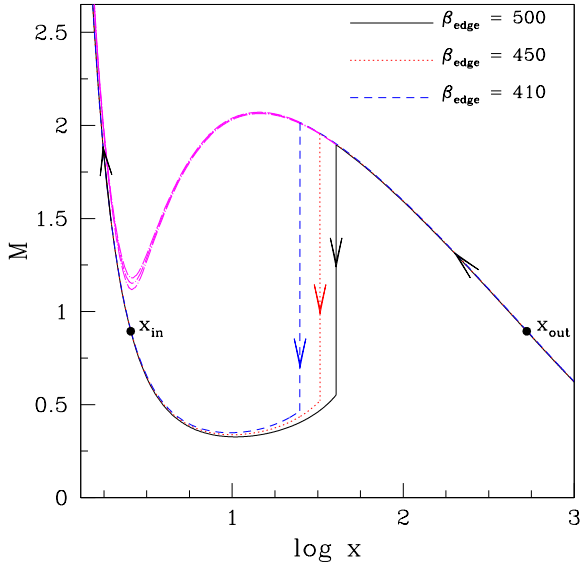}
\end{center}
\caption{Plot of Mach number with logarithmic radial distance for different values
of $\beta_{\rm edge}$. Accreting flows are injected from $x_{\rm edge} = 1000$ with
$\lambda_{\rm edge} = 1.886$, ${\cal E}_{\rm edge}=1.9133\times10^{-4}$, $\alpha_B = 0.01$
and $\xi=20$. Solutions represented by the solid, dotted and dashed curves are
for $\beta_{\rm edge}=500$, $450$ and $410$ respectively. The corresponding shock
locations are indicated by the vertical arrows as $x_s = 40.60$ (solid), $32.69$
(dotted) and $24.94$ (dashed). Sonic points are marked by the filled circles.
See text for details.
}
\end{figure}

In our subsequent analysis, 
we explore the response of $\beta_{\rm edge}$ on shock dynamics.
While doing this, we inject matter from the outer edge at $x_{\rm edge} = 1000$ with
$\lambda_{\rm edge} = 1.886$, ${\cal E}_{\rm edge}=1.9133\times10^{-4}$, $\alpha_B = 0.01$
and $\xi=20$, and vary $\beta_{\rm edge}$. In Fig. 5, solid, dotted and dashed 
curves represent the results corresponding to $\beta_{\rm edge} = 500, 450$ and $410$,
respectively. Here, the shock front moves inward as $\beta_{\rm edge}$ is decreased. This 
eventually indicates the fact that the size of the post-shock corona decreases with
the increase of the magnetic pressure inside the disc. In reality, the decrease
of $\beta_{\rm edge}$ implies the increment of magnetic turbulence inside the disc.
The growth of the turbulent magnetic field increases of Maxwell stress
that leads to enhance the angular momentum transport outward. Hence, the
centrifugal support against gravity becomes weak that pushes shock front inward.
The dynamics of the shock location is eventually controlled due to the combined
effect of cooling and magnetic field.

\begin{figure}
\begin{center}
\includegraphics[width=0.45\textwidth]{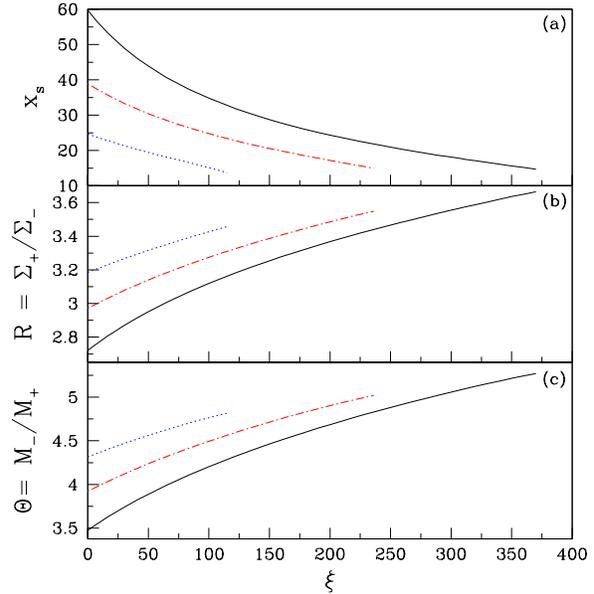}
\end{center}
\caption{Variation of different flow properties such as (a) the shock location $x_s$,
(b) shock compression ratio $R$, and (c) shock strength $\Theta$ as a function of
$\xi$ for flows injected from $x_{\rm edge} = 1000$ with $\alpha_B = 0.01$,
$\beta_{\rm edge} = 550$ and ${\cal E}_{\rm edge}=1.9133 \times 10^{-4}$ but with different
$\lambda_{\rm edge}$. Solid curve corresponds to the result obtained for
$\lambda_{\rm edge} = 1.890$ and the dot-dashed and dotted curves are for
$\lambda_{\rm edge} = 1.873$ and $1.856$, respectively. See text for details.
}
\end{figure}

In Fig. 6, we present the comparison of shock properties as function of the cooling
efficiency factor ($\xi$). In the upper panel (Fig. 6a), we show the variation of 
shock locations for different values of $\lambda_{\rm edge}$.
Here, we choose the outer edge of the disc at $x_{\rm edge} = 1000$ and inject matter
with ${\cal E}_{\rm edge}=1.9133\times 10^{-4}$, $\beta_{\rm edge} = 550$ and $\alpha_B = 0.01$
for all cases. The solid curve denotes the result corresponding to $\lambda_{\rm edge} = 1.890$
and the dot-dashed and dotted curves are for $\lambda_{\rm edge} = 1.873$ and $1.856$, respectively. 
It is clear from the figure that stationary shocks in an accretion flow can be
obtained for a wide range of $\xi$. For a given $\lambda_{\rm edge}$,
the shock front is shifted towards the horizon
with the increase of the cooling factor ($\xi$) as depicted in Fig. 4. This is
because the flow loses its energy due to cooling during accretion. With this, when
$\xi$ exceeds its critical value, shock disappears as the standing shock
conditions are not satisfied. This eventually provides an indication that the possibility
of stationary shock transition is likely to be reduced with the increase of $\xi$.
Evidently, the critical value of $\xi$ largely depends on the
accretion flow parameters at the outer edge. Moreover, above the critical
cooling limit, the accretion flow still may contain shock waves which are oscillatory
in nature and the investigation of such shock properties is beyond the scope of the present
paper. In addition, for a given $\xi$, shock recedes away from the horizon
when $\lambda_{\rm edge}$ is increased. 
This is not surprising as the large
$\lambda_{\rm edge}$ enhances the strength of the centrifugal barrier that pushes the
shock front outside. This clearly indicates that the centrifugal force seems to play a
crucial role in deciding the possibility of shock formation.

As discussed in Section 2 that the bremsstrahlung emissivity directly depends on the
density and temperature of the flow and therefore, the emergent radiations
from the disc are also depend on them. Hence, it is useful to calculate the density
and temperature distributions of the flow across the shock discontinuity as both the density
and temperature are enhanced due to shock compression in the post-shock flow.
For that, first we calculate the compression ratio that determines the density compression
of the flow across the shock and is defined as the ratio of the vertically averaged post-shock
density to the pre-shock density ($R=\Sigma_{+}/\Sigma_{-}$). In Fig. 6b, we plot the
variation of compression ratio as function of cooling efficiency factor for the same set
of input parameters as in Fig. 6a. A positive correlation
is observed in all cases as the compression ratio is increased with the increase of
cooling rate. This is quite natural because higher cooling efficiency
pushes the shock front inward that causes more compression in the post-shock flow and
eventually, compression ratio increases.
When the cooling efficiency factor is reached its critical value, we observe a cut-off in
the variation of compression ratio. This happens in all cases as the standing shock
fails to exist there. Further, we calculate the shock strength ($\Theta$) which is
defined as the ratio of pre-shock Mach number ($M_{-}$) to the post-shock Mach number
($M_{+}$) and it measures the temperature jump across the
shock. In Fig. 6c, we plot shock strength as function of $\xi$ for the same
set of input parameters as in Fig. 6a and observe the variation of $\Theta$ 
very similar to $R$ as depicted in Fig. 6b.

\begin{figure}
\begin{center}
\includegraphics[width=0.45\textwidth]{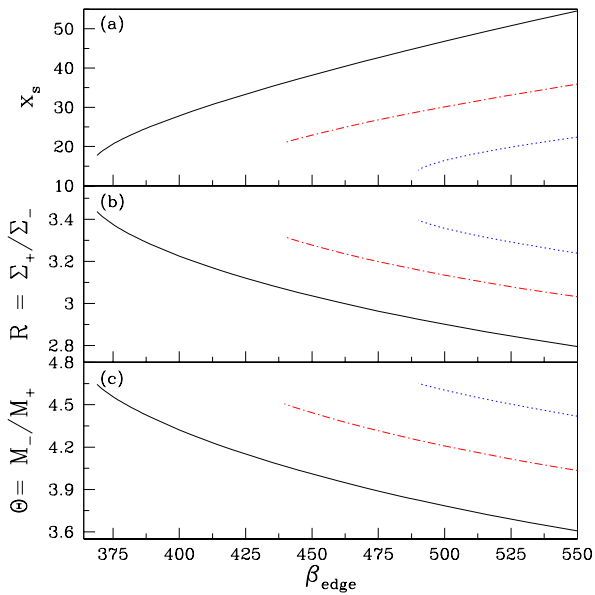}
\end{center}
\caption{Variation of different flow properties such as (a) the shock location $x_s$,
(b) shock compression ratio $R$, and (c) shock strength $\Theta$ as a function of
$\beta_{\rm edge}$ for different values of $\lambda_{\rm edge}$. Flows is injected from
$x_{\rm edge} = 1000$ with ${\cal E}_{\rm edge}=1.9133\times 10^{-4}$, $\alpha_B = 0.01$ and
$\xi=20$, respectively. Results represented by the solid curve is for
$\lambda_{\rm edge}=1.892$ and the dot-dashed and dotted curves are for $\lambda_{\rm edge}=1.874$
and $1.856$. See text for details.
}
\end{figure}

We continue our study to investigate the shock properties in terms of 
$\beta_{\rm edge}$ for flows with same outer boundary values, namely, $x_{\rm edge} = 1000$,
${\cal E}_{\rm edge}=1.9133\times 10^{-4}$, $\alpha_B = 0.01$ and $\xi=20$. The solid,
dot-dashed, dotted curves are for $\lambda_{\rm edge} = 1.892, 1.874$ and $1.856$, respectively.
As discussed in
Fig. 5, here also the shock location is reduced with the decrease of $\beta_{\rm edge}$
for all cases having different angular momentum at the outer edge. Interestingly, 
the lower limit of $\beta_{\rm edge}$ is not indefinite, because the possibility of shock
transition ceases to exist when $\beta_{\rm edge}$ is reduced to its critical limit.
With this, we compute the shock compression ratio ($R$) and the shock strength
($\Theta$) as in Fig. 6(b-c) and find that both are increased when $\beta_{\rm edge}$
is decreased gradually.

\begin{figure}
\begin{center}
\includegraphics[width=0.45\textwidth]{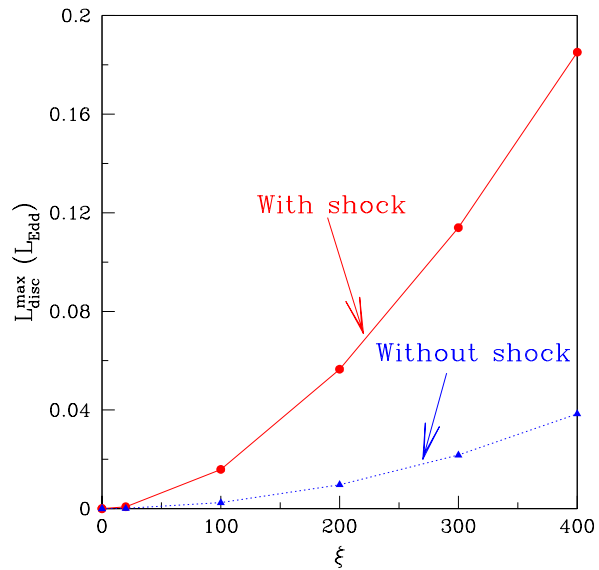}
\end{center}
\caption{Variation of maximum disc luminosity $L^{\rm max}_{\rm disc}$ as a function of cooling
efficiency factor $\xi$. Filled circles connected with solid line and filled triangles
connected with dotted line are for shocked and shock free accretion solutions.
}
\end{figure}


\subsection{Accretion Disk Luminosity}

In this work, we consider the Bremsstrahlung emission process as the prospective
cooling mechanism for flows accreting on to black holes. Following this, we estimate
the disc luminosity ($L_{\rm disc}$) as,

$$
L_{\rm disc} = 4\pi \int_{x_{\rm in}}^{x_{\rm edge}} Q^{-}x dx
$$
where, $x_{\rm in}$ and $x_{\rm edge}$ denote the inner sonic point and the outer edge of the
disc, respectively and $Q^{-}$ is the Bremsstrahlung cooling rate. Here, we neglect
radiations emitted from the region between the horizon and the inner sonic point as they are
expected to be red-shifted and do not contribute significantly in the disc luminosity.
In Fig. 8, we present the variation of maximum Bremsstrahlung luminosity as function of
cooling efficiency factor ($\xi$). Filled circles connected with solid lines denote
the results obtained from the shock induced global accretion solution whereas the
filled triangles joined with dotted lines represent the results for shock free
accretion solutions. For a given cooling efficiency factor,
we compute the maximum disc luminosity employing our model for shock and shock free cases.
In general, we observe that the total luminosity is enhanced when $\xi$ is increased.
This is because the rise of $\xi$ essentially increase the density of the flow and 
consequently flow cools efficiently. In addition, we find that for a given $\xi$,
the disc luminosity is always higher for flows containing shock waves compared to the
flows having no shocks. This apparently provides an indication that the shocked accretion
solutions are perhaps potentially more preferred to study the energetics of the black
hole sources.

\begin{figure}
\begin{center}
\includegraphics[width=0.45\textwidth]{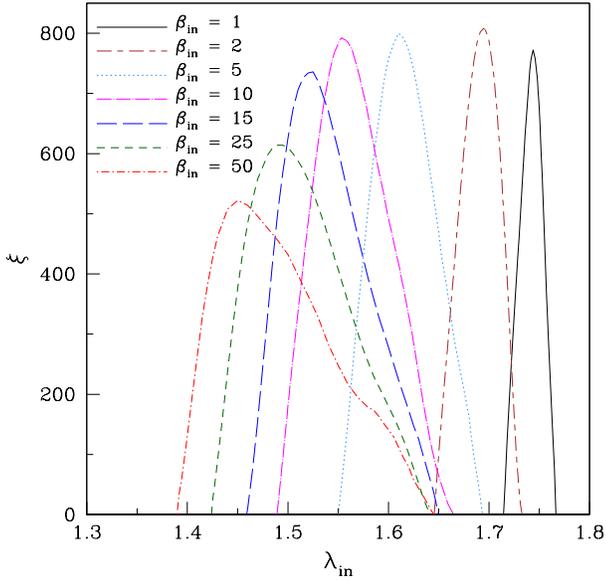}
\end{center}
\caption{Separations of the parameter space that allow stationary shock waves in the
$\lambda_{\rm in}-{\xi}$ plane. Dot-dashed, dashed, long-dashed, dot-long dashed, dotted,
short-long dashed and solid curves are for $\beta_{\rm in}= 50, 25, 15, 10, 5, 2$,
and $1$, respectively. Here, we fix $\alpha_B=0.01$. See text for details.
}
\end{figure}
\begin{figure}
\begin{center}
\includegraphics[width=0.45\textwidth]{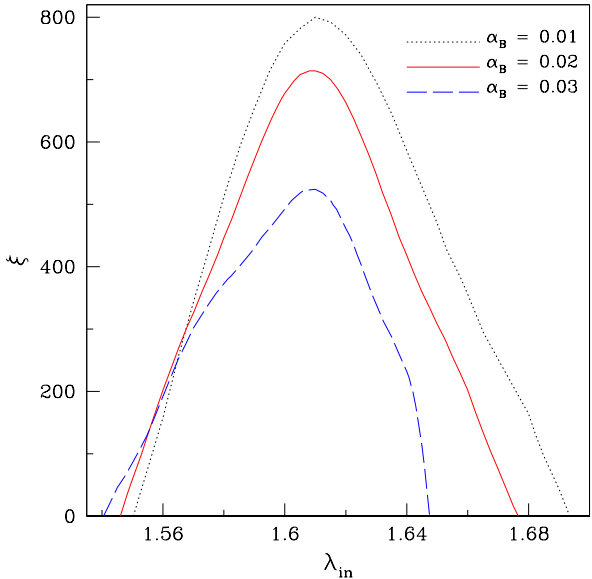}
\end{center}
\caption{Effective regions of the parameter space for stationary shock in the
$\lambda_{\rm in}-\xi$ plane. The regions separated by dotted, solid and dashed
curves are for $\alpha_{B}=0.01, 0.02$, and $0.03$, respectively. Here, we fix
$\beta_{\rm in}=5$. See text for details.
}
\end{figure}

\subsection{Parameter Space for Shock}

It is already pointed out that
the dissipative global accretion solutions including
shock waves are not the isolated solutions, instead such solutions exist
for a wide range of angular momentum and the cooling efficiency factor.
In order to understand the influence of magnetic field on the properties of
the stationary shock waves in a dissipative accretion flow, we
identify the region of the parameter space spanned by the angular momentum at
the inner sonic point ($\lambda_{\rm in}$) and the cooling efficiency factor ($\xi$)
that provides shock solutions and subsequently classify them in terms of
$\beta_{\rm in}$. Here, $\beta_{\rm in}$ refers to the value of $\beta$ measured at
the inner sonic point $x_{\rm in}$. The results are depicted in Fig. 9 where, we
choose $\alpha_{B}=0.01$.
The dot-dashed boundary separates the shock parameter space and is obtained for
$\beta_{\rm in}=50$ where magnetic pressure is weak and accretion flow is tended to be
gas pressure dominated.
As the strength of the magnetic pressure is increased relative to the gas pressure,
the parameter space shifts towards the higher angular momentum side. This is due
to the fact that the range of angular momentum at the inner sonic point for transonic
accretion flow increases when $\beta_{\rm in}$ is decreased. Here, dashed, long-dashed,
dot-long dashed, dotted, short-long dashed and solid curves identify the boundary
for  $\beta_{\rm in}=25, 15, 10, 5$ $2$ and $1$, respectively.
We observe that when the accretion flow starts dominated by the magnetic pressure, the
effective region of the parameter space for standing shocks reduces gradually and
finally disappears when $\beta_{\rm in}$ reached its critical value.

We continue our study of parameter space to explore the role of viscous dissipation in the
shock parameter space. While doing so, we choose $\beta_{in}=5$ all throughout 
and obtain the parameter space as function of $\alpha_{B}$ which is depicted in Fig. 10.
As before, here again we find that shock induced global accretion solutions can be obtained
for a wide
range of input parameters, namely $\lambda_{in}$ and $\xi$. In the figure, the
viscous dissipation parameters are marked. We observe that as the dissipation is increased,
the parameter space for stationary shock is shrunk. This is simply because the possibility
of shock transition is reduced with the enhancement of dissipation in the flow.
Eventually, the shock parameter space disappears when $\alpha_{B}$ is crossed
its critical value.

\subsection{Critical Viscosity Parameter}

\begin{figure}
\begin{center}
\includegraphics[width=0.45\textwidth]{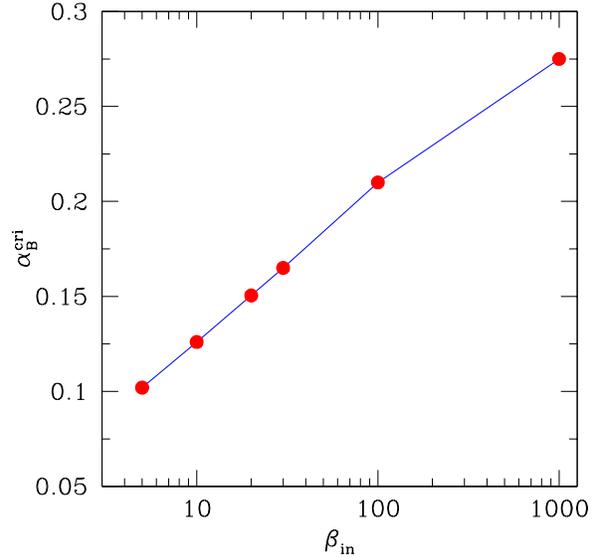}
\end{center}
\caption{Variation of critical viscosity parameter ($\alpha_{B}^{\rm cri}$) with
$\beta_{\rm in}$ that allows standing shocks. Here, we consider $\xi=20$.
See text for details.
}
\end{figure}


In the previous Section, we have pointed out that the dynamical structure of the 
global accretion flow changes when the viscosity parameter exceeds its critical value.
Following this, we obtain the value of the critical viscosity parameter
$\alpha_{B}^{\rm cri}$ based on the criteria of whether a standing shock is formed
or not. Evidently, the critical viscosity parameter largely depends on the inflow
parameters. In Fig. 11, we demonstrate the variation of $\alpha_{B}^{\rm cri}$
with $\beta_{\rm in}$ for $\xi=20$. In a magnetized flow, the angular momentum transport
in the disc equatorial plane is increased as the magnetic pressure contributes to the
total pressure. Hence, a lower value of $\alpha_{B}$ is sufficient to transport 
angular momentum required for shock formation. On the contrary, the possibility of
shock formation is enhanced with the higher viscosity parameter when the flow is shifted
towards the gas pressure dominated regime. As $\beta_{\rm in}$ is increased, the
critical viscosity parameter $\alpha_{B}^{\rm cri}$ tends to approach
$\alpha_{\Pi}^{\rm cri}$ ($\sim 0.3$) as estimated by the \citet{Chakrabarti-Das04}
for gas pressure dominated flow.

\section{Astrophysical Applications}
\begin{center}
\begin{table*}
\begin{minipage}{180mm}
\caption{Estimation of shock luminosity. Column 1 lists the names of the sources and
Column 2-3 give mass and accretion rate. Column 4-7 provide the model parameters
and Column 8-10 denote the maximum energy dissipation, shock location and estimated
maximum shock luminosity obtained from Eq. (20). Observed core radio luminosity values are given
in column 11.}
\label{my-label}
\begin{tabular}{l c c c c c c c c c c}
\hline\hline\\
Object & $M_{BH}$ & $\dot{m}$ & $\mathcal{E}_{\rm out}$ & $\lambda_{\rm out}$ &
$\beta_{\rm out}$ & $\Delta \mathcal{E}^{'}$ &  $\Delta \mathcal{E}^{\rm max}$ & $x_s$ &
$L^{\rm max}_{\rm shock}$ & $L^{\rm Obs}_{\rm jet}$\\

&$(M_{\odot})$ & $({\dot M}_{Edd})$ & $(10^{-4} c^2)$ & $(cr_g)$ & &  & ($10^{-3} c^2$) & $(r_g)$ & (erg/s) & (erg/s) \\ \hline\\
 Sgr A*   & $4.90 \times 10^{6}$$~^a$ &  $2.776 \times 10^{-4}$$~^b$  & $2.356 $ & $1.643$ & $1000$ & $0.65$ &
$3.637$ &  $13.66$ &  $6.2 \times 10^{38}$ & $1.0 \times 10^{39}$$~^c$ \\
 NGC 4258 & $3.39\times 10^{7}$$~^d$ & $9.423 \times 10^{-5}$$~^e$ & $1.996$ & $1.637$ & $ 3500$ & $0.71$ &
$3.937$ & $14.48$ & $1.6 \times 10^{39}$ & $1.0 \times10^{42}$$~^c$ \\
NGC 3079 & $6.76\times 10^{7}$$~^d$ & $5.907 \times 10^{-3}$$~^e$ & $2.129$ & $1.630$ & $ 2400$ & $0.64$ &
$3.835$ & $13.42$ & $1.9 \times 10^{41}$ & $ 4.0 \times 10^{41}$$~^f$\\
 Mrk 79   & $5.24\times 10^{7} $$~^g$ & $1.125 \times 10^{-2} $$~^h$ & $2.219 $ & $1.635$ & $2000$ & $0.74$ &
$4.242 $ &  $13.39$  & $3.1 \times 10^{41}$ & $3.4 \times 10^{40}$$~^h$ \\
NGC 6500 & $1.70\times 10^{8}$$~^d$ & $8.000 \times 10^{-6} $$~^i$ & $2.125$ & $1.633$ & $ 3000$ & $0.70$ &
$4.089$ & $13.51$ & $6.9 \times 10^{38}$ & $7.9 \times 10^{38}$$~^j$\\
  M87     & $3.50\times 10^{9}$$~^k$ & $1.192 \times 10^{-4} $$~^l$ &  $2.468$ & $1.669$ & $330$ & $0.34$  &
$1.938  $ &   $13.72$ &  $1.0 \times 10^{41}$ & $5.0 \times 10^{44}$$~^m$ \\
\hline
\end{tabular}\\
References: $^a$\citet{Aschenbach10}, $^b$\citet{Yuan-etal02}, $^c$\citet{Falcke-Biermann99},
$^d$\citet{Kadowaki-etal15}, $^e$\citet{Yamauchi-etal04}, $^f$\citet{Shafi-etal15},
$^g$\citet{Peterson-etal04}, $^h$\citet{Riffel-etal13}, $^i$\citet{Satyapal-etal04},
$^j$\citet{Falcke-etal04}, $^k$\citet{Walsh-etal13}, $^l$\citet{Kuo-etal14},
$^m$\citet{deGasperin-etal12}.
\end{minipage}
\end{table*}
\end{center}

So far, we have concentrated on the accretion shocks around black holes where the specific
energy across the shock front is considered to be constant \citep{Chakrabarti89} and
these shocks 
are radiatively inefficient in nature. However, in reality, the characteristic
of the shocks can be dissipative as well where a part of the accreting energy is released
vertically through the disc surface at the shock location causing the reduction of
specific energy in the PSC \citep{Singh-Chakrabarti11}. Usually the energy
dissipation mechanism at the shock is regulated by the thermal
Comptonization process \citep{Chakrabarti-Titarchuk95} and therefore, the thermal
distribution in the PSC is reduced. Based on this criteria, we estimate the energy
loss across the shock where we assume that the loss of energy is 
scaled with the temperature difference
between the immediate post-shock and pre-shock flow and is given by \citep{Das-etal10},
$$
\Delta {\cal E}=\Delta {\cal E}^{'}(a_{+}^2-a_{-}^2),
\eqno(19)
$$
where, $a_{+}$ and $a_{-}$ are the post-shock and pre-shock sound speeds,
respectively and $\Delta {\cal E}^{'}$ denotes the fraction of the thermal energy
difference lost in this process
which we treat as a parameter. For a weakly rotating black hole, Das et al, 2010
calculated the maximum
energy dissipation at the shock and is estimated as $\Delta {\cal E}^{\rm max} \sim 2.5 \%$.
Needless to mentioned that
$\Delta {\cal E}$ chosen beyond this range does not provide global transonic 
accretion solution including shock waves.

In this scenario, the accessible energy at the PSC is same as the 
available energy dissipated at the shock. A fraction of this energy is converted
in to high energy radiations and the remaining part of the energy is utilized to
produce jets as they are likely to originate from the PSC around
the black holes. Subsequently, these jets simultaneously ingest a part of this
energy for the work done against gravity and for carrying out their thermodynamical
expansion. 
The remaining part of the energy is then utilized to power the jets. Therefore,
according to the energy budget, the total usable energy available in the post-shock
flow is $\Delta {\cal E}$ and the corresponding loss of kinetic power from the
disc can be estimated in terms of the observable quantities as in
\citet{Le-Becker04,Le-Becker05},
$$
L_{\rm total}=L_{\rm shock}=\dot M \times \Delta {\cal E} \times c^2~~{\rm erg~s^{-1}},
\eqno(20)
$$
where, $L_{\rm total}$ is the kinetic power lost by the disc, $L_{\rm shock}$ is the shock
luminosity and $\dot M$ is the accretion rate for a given source, respectively.
Following the above approach, we estimate the maximum shock luminosity
($L^{\rm max}_{\rm shock}$) that corresponds to maximum energy dissipation at the shock.
Here, $\alpha_B=0.001$ and $\xi = 10$ are considered for all cases.
In Table 1, we present the physical parameters of the super-massive black hole 
sources including model parameters and estimated maximum shock luminosity.
In column 1-3, we display the list of sources, their mass ($M_{BH}$) and
dimensionless accretion rate (${\dot m}$). In column 4-6, we indicate the representative values
of flow variables at the outer sonic point, namely energy ${\mathcal E}_{\rm out}$,
angular momentum
$\lambda_{\rm out}$ and $\beta_{\rm out}$. In column 7, we mention the $\Delta {\mathcal E}^{'}$
value that provides the maximum $\Delta {\mathcal E}^{\rm max}$ (in column 8) and shock location
$x_s$ (in column 9). In column 10, we present the maximum shock luminosity 
$L^{\rm max}_{\rm shock}$. In this analysis, our motivation is to quantify the upper
limit of the energy that can be extracted from the PSC to power the deflected matter from
the disc as Jets.
Therefore, we calculate the maximum energy dissipation $\Delta {\mathcal E}^{\rm max}$
at the location of shock transition. We find that the estimated shock luminosities
for super-massive black hole sources under consideration are in close agreement
with the observed core radio luminosity values $L^{\rm Obs}_{\rm jet}$ (in column 11)
\citep{Falcke-Biermann99,Falcke-etal04,deGasperin-etal12,Riffel-etal13,Shafi-etal15}.

\section{Conclusions}

In this paper, we have studied the dynamical structure of a magnetized
accretion flow around a non-rotating black hole in presence of Bremsstrahlung cooling.
Since the exact physical mechanism for angular momentum transport 
in an accretion disc is not yet conclusive, we assume that the Maxwell
stress is proportional to the total pressure following the work of \cite{Machida-etal06},
where the constant of proportionality $\alpha_B$ plays the role similar to the
conventional viscosity parameter as described in \cite{Shakura-Sunyaev73}.
We indeed find that such an accretion flow is transonic in nature. This is because
the inflowing matter must satisfy the inner boundary condition imposed by the black
hole horizon. Depending on the flow parameters, namely angular momentum ($\lambda$),
viscosity ($\alpha_B$), cooling efficiency factor ($\xi$) and $\beta$
respectively, accreting matter changes its sonic state multiple times as
it contains multiple sonic points. Flows of this kind are of special interest as
they may contain shock wave
which is perhaps essential to understand the spectral and timing properties of
the black hole candidates \citep{Chakrabarti-Manickam00,Nandi-etal01a,Nandi-etal01b,
Nandi-etal12,Radhika-Nandi14,Iyer-etal15}.

In Section 3, we calculate the shock induced global accretion solution in presence
of toroidal magnetic field. Due to shock transition, the post-shock flow, \eg~PSC
is compressed and as a consequence PSC becomes hot and dense as is seen in Fig. 3.
According to our solutions, PSC remains optically thin though there is a sharp rise of
density at the inner part of the disc. This effectively enhances the possibility of
escaping the hard radiations from PSC. When the cooling efficiency is increased,
the thermal pressure of PSC is evidently reduced.
As a consequence, shock front moves towards the horizon and finally settles
down at a smaller radius where total pressure across the shock front is balanced.
Above the critical cooling limit ($\xi^{\rm cri}$), PSC disappears
due to effect of excess cooling where shock conditions are not favorable. It must be
noted that $\xi^{\rm cri}$ does not correspond to a unique value as
it depends of the other flow parameters.

One of the important results of this work is to obtain the global shock
solutions in gas pressure dominated flow as well as magnetic pressure dominated
flow and subsequently investigate the dependencies of flow parameters on shock
properties. In Fig. 6-7, we observe that global shock solutions are not the
isolated solutions, instead shock may form for a wide range of flow parameters.
Moreover, we find that $\alpha_B$ and ${\beta}$ play important role in deciding
the formation of shock waves (Fig. 9-10).

We also calculate the critical viscosity parameter $(\alpha^{\rm cri}_{B})$ that
allows standing shocks in the accretion flow around black holes. Beyond this critical limit,
standing shock conditions are not favorable and hence, steady shock ceases to exist.
We find that $\alpha^{\rm cri}_{B}$ gradually increases as the plasma
$\beta$ increases and ultimately tends to
the value $\sim 0.3$ as reported by \citet{Chakrabarti-Das04} for gas
pressure dominated flow (Fig. 11). For $\alpha_{B} > \alpha^{\rm cri}_{B}$,
however, oscillatory shocks may still form \citep{Das-etal14} which is the next issue
to be undertaken and will be reported elsewhere.

Further, we self-consistently study the characteristics of the dissipative shock
solutions. In this scenario, a part of the accreting energy is escaped
from the shock location in the vertical directions through the disc surface
and this dissipated energy is being utilized to power the jets
\citep{Chakrabarti-Titarchuk95,Le-Becker04,Le-Becker05}. 
In order to understand the implications of the dissipative shock, we estimate
the maximum shock luminosity ($L^{\rm max}_{\rm shock}$) corresponding to the maximum
energy dissipation ($\Delta {\cal E}^{\rm max}$) at the shock using equations (19) and (20).
In Table 1, we summarize the physical parameters of the black hole sources along
with the model parameters and $L^{\rm max}_{\rm shock}$. We observe that the
estimated $L^{\rm max}_{\rm shock}$ for several super-massive black hole sources are
in close agreement with the observed core radio luminosity values ($L^{\rm Obj}_{\rm jet}$).

Finally, we point out that the present formalism is developed based on some
approximations. We ignore the rotation of the black hole and use pseudo-Newtonian
potential to describe the space-time geometry around a non-rotating
black hole as it allows us to study the non-linear shock solutions in a simpler way.
We neglect the synchrotron emission process in this work although it is
expected to play an role in a magnetized accretion flow. An extension of our present
study including synchrotron cooling to the case of rotating black hole is under
progress and will be reported elsewhere. Also, the adiabatic index of the flow is
considered to be a global constant instead of calculating it self-consistently
using thermal properties of the flow. Of course, the implementation
of such issue is beyond the scope of the present paper, however, we believe that the
basic conclusions of this work will remain qualitatively unaltered.

\section*{Acknowledgments}
Authors would like to thank Anuj Nandi for discussions. Authors also thank
the anonymous referee for useful comments and constructive suggestions.



\begin{thebibliography}{reference}

\bibitem[\protect\citeauthoryear{Akizuki \& Fukue}{2006}]{Akizuki-Fukue06}
Akizuki, C., Fukue, J., 2006, PASJ, 58, 469

\bibitem[\protect\citeauthoryear{Aktar \etal}{2015}]{Aktar-etal15}
Aktar R., Das S., Nandi A., 2015, \mnras, 453, 3414

\bibitem[\protect\citeauthoryear{Aschenbach \etal}{2010}]{Aschenbach10} 
Aschenbach B., 2010, \MmSAI, 81, 319

\bibitem[\protect\citeauthoryear{Balbus \& Hawley}{1991}]{Balbus-Hawley91}
Balbus, S., \& Hawley, J. F. 1991, \apj, 376, 214

\bibitem[\protect\citeauthoryear{Balbus \& Hawley}{1998}]{Balbus-Hawley98}
Balbus, S., \& Hawley, J. F. 1998, RvMP, 70, 1

\bibitem[\protect\citeauthoryear{Becker \& Kazanas}{2001}]{Becker-Kazanas01}
Becker P. A., Kazanas D., 2001, \apj, 546, 429

\bibitem[\protect\citeauthoryear{Becker \etal}{2008}]{Becker-etal08}
Becker P. A., Das S., Le T., 2008, \apjl, 677, 93

\bibitem[\protect\citeauthoryear{Begelman \& Pringle}{2007}]{Begelman-Pringle07}
Begelman, M. C., Pringle, J.E., 2007, \mnras, 375, 1070

\bibitem[\protect\citeauthoryear{Bu \etal}{2009}]{Bu-etal09}
Bu, D.-F., Yuan, F., \& Xie, F.-G., 2009, \mnras, 392, 325

\bibitem[\protect\citeauthoryear{Chakrabarti}{1989}]{Chakrabarti89}
Chakrabarti, S. K., 1989, \apj, 347, 365

\bibitem[\protect\citeauthoryear{Chakrabarti}{1990}]{Chakrabarti90}
Chakrabarti, S. K., 1990, `Theory of Transonic Astrophysical 
Flows' by Sandip K. Chakrabarti, (Singapore, World Scientific 
Publishing Co. Ltd.).

\bibitem[\protect\citeauthoryear{Chakrabarti \& Titarchuk}{1995}]{Chakrabarti-Titarchuk95} 
Chakrabarti, S K., Titarchuk, L., 1995, \apj, 455, 623.

\bibitem[\protect\citeauthoryear{Chakrabarti}{1996}]{Chakrabarti96}
Chakrabarti, S. K., 1996, \apj, 464, 664

\bibitem[\protect\citeauthoryear{Chakrabarti}{1999}]{Chakrabarti99}
Chakrabarti S. K., 1999, \aap, 351, 185

\bibitem[\protect\citeauthoryear{Chakrabarti \& Manickam}{2000}]{Chakrabarti-Manickam00}
Chakrabarti, S. K., Manickam, S. G., 2000, \apj, 531, L41

\bibitem[\protect\citeauthoryear{Chakrabarti \& Das}{2004}]{Chakrabarti-Das04}
Chakrabarti, S. K., Das, S., 2004, \mnras, 349, 649

\bibitem[\protect\citeauthoryear{Chattopadhyay \& Chakrabarti}{2000}]
{Chattopadhyay-Chakrabarti00} Chattopadhyay I., Chakrabarti S. K., 2000, \ijmpd, 9, 717

\bibitem[\protect\citeauthoryear{Chattopadhyay \& Chakrabarti}{2002}]
{Chattopadhyay-Chakrabarti02}Chattopadhyay, I., Chakrabarti, S. K., 2002, \mnras, 333, 454

\bibitem[\protect\citeauthoryear{Chattopadhyay \& Das}{2007}]{Chattopadhyay-Das07}
Chattopadhyay, I., Das, S., 2007, \na, 12, 454

\bibitem[\protect\citeauthoryear{Das \etal}{2001a}]{Das-etal01a}
Das S., Chattopadhyay I., Chakrabarti S. K., 2001a, \apj, 557, 983

\bibitem[\protect\citeauthoryear{Das \etal}{2001b}]{Das-etal01b}
Das S., \etal, 2001b, \aap, 379, 683

\bibitem[\protect\citeauthoryear{Das \& Chakrabarti}{2004}]{Das-Chakrabarti04}
Das S., Chakrabarti S. K., 2004, \ijmpd, 13, 1955

\bibitem[\protect\citeauthoryear{Das}{2007}]{Das07} 
Das, S., 2007, MNRAS, 376, 1659

\bibitem[\protect\citeauthoryear{Das \& Chattopadhyay}{2008}]{Das-Chattopadhyay08} 
Das, S., Chattopadhyay I., 2008, New Atsron., 13, 549

\bibitem[\protect\citeauthoryear{Das \etal}{2009}]{Das-etal09}
Das S., Becker P. A., Le T., 2009, \apj, 702, 649

\bibitem[\protect\citeauthoryear{Das \etal}{2010}]{Das-etal10}
Das S., Chakrabarti S. K., Mondal S., 2010, \mnras, 401, 2053

\bibitem[\protect\citeauthoryear{Das \etal}{2014}]{Das-etal14}
Das S., \etal, 2014, \mnras, 442, 251

\bibitem[\protect\citeauthoryear{de Gasperin \etal}{2012}]{deGasperin-etal12}
de Gasperin F., \etal, 2012, \aap, 547, 56

\bibitem[\protect\citeauthoryear{Falcke \& Biermann}{1999}]{Falcke-Biermann99}
Falcke H., Biermann P. L., 1999, \aap, 342, 49

\bibitem[\protect\citeauthoryear{Falcke \etal}{2004}]{Falcke-etal04}
Falcke H., Kording* E., Markoff S., 2004, A\&A, 414, 895

\bibitem[\protect\citeauthoryear{Fukue}{1987}]{Fukue87}
Fukue J., 1987, PASJ, 39, 309

\bibitem[\protect\citeauthoryear{Gierli\'nski \& Newton}{2006}]{Gierlinski-Newton06}
Gierli\'nski M., Newton J., 2006, \mnras, 370, 837

\bibitem[\protect\citeauthoryear{Gu \& Lu}{2004}]{Gu-Lu04}
Gu W. M., Lu J. F., 2004, ChPhL, 21, 2551

\bibitem[\protect\citeauthoryear{Hirose \etal}{2006}]{Hirose-etal06}
Hirose S., Krolik L. H., Stone J. M., 2006, \apj, 640, 901

\bibitem[\protect\citeauthoryear{Iyer \etal}{2015}]{Iyer-etal15}
Iyer N., Nandi A., Mandal S., 2015, \apj, 807, 108

\bibitem[\protect\citeauthoryear{Kadowaki \etal}{2015}]{Kadowaki-etal15}
Kadowaki L. H. S., de Gouveia Dal Pino E. M., Singh C. B., 2015, \apj, 802, 113

\bibitem[\protect\citeauthoryear{Kuo \etal}{2014}]{Kuo-etal14}
Kuo C. Y., \etal, 2014, \apj, 783, 33

\bibitem[\protect\citeauthoryear{Landau \& Lifshitz}{1959}]{Landau-Lifshitz59}
Landau L. D., Lifshitz, E. D., 1959, Fluid  Mechanics (New York: Pergamon)

\bibitem[\protect\citeauthoryear{Le \& Becker}{2004}]{Le-Becker04}
Le T., Becker P. A., 2004, \apjl, 617, 25

\bibitem[\protect\citeauthoryear{Le \& Becker}{2005}]{Le-Becker05}
Le T., Becker P. A., 2005, \apjl, 632, 476

\bibitem[\protect\citeauthoryear{Lu \etal }{1999}]{Lu-etal99}
Lu J. F., Gu W. M., Yuan F., 1999, ApJ, 523, 340

\bibitem[\protect\citeauthoryear{Machida \etal}{2006}]{Machida-etal06}
Machida, M., Nakamura, K. E. \& Matsumoto, R., 2006, \pasj, 58, 193.

\bibitem[\protect\citeauthoryear{Matsumoto \etal}{1984}]{Matsumoto-etal84}
Matsumoto, R., Kato, S., Fukue, J. \& Okazaki, A. T., 1984, \pasj, 36, 71

\bibitem[\protect\citeauthoryear{Molteni \etal}{1994}]{Molteni-etal94}
Molteni D., Lanzafame G., Chakrabarti S. K., 1994, \apj, 425, 161

\bibitem[\protect\citeauthoryear{Molteni \etal}{1996}]{Molteni-etal96}
Molteni D., Ryu D., Chakrabarti S. K., 1996, \apj, 470, 460

\bibitem[\protect\citeauthoryear{Nagakura \& Yamada}{2009}]{Nagakura-Yamada09}
Nagakura H., Yamada S., 2009, \apj, 696, 2026 

\bibitem[\protect\citeauthoryear{Narayan \etal}{1997}]{Narayan-etal97} 
Narayan, R., Kato, S. \& Honma, F., 1997, \apj, 476, 49.

\bibitem[\protect\citeauthoryear{Nandi \etal}{2001a}]{Nandi-etal01a}
Nandi A., \etal, 2001a, \aap, 380, 245

\bibitem[\protect\citeauthoryear{Nandi \etal}{2001b}]{Nandi-etal01b}
Nandi A., \etal, 2001b, \mnras, 324, 267

\bibitem[\protect\citeauthoryear{Nandi \etal}{2012}]{Nandi-etal12}
Nandi A., Debnath D., Mandal S., Chakrabarti S. K., 2012, \aap, 542, 56

\bibitem[\protect\citeauthoryear{Okuda}{2014}]{Okuda14}
Okuda T., 2014, \mnras, 441, 2354

\bibitem[\protect\citeauthoryear{Okuda \& Das}{2015}]{Okuda-Das15}
Okuda T., Das S., 2015, \mnras, 453, 147

\bibitem[\protect\citeauthoryear{Oda \etal}{2007}]{Oda-etal07} 
Oda, H., Machida, M., Nakamura, K. E. \& Matsumoto, R., 2007, \pasj, 59, 457

\bibitem[\protect\citeauthoryear{Oda \etal}{2010}]{Oda-etal10} 
Oda H., Machida M., Nakamura K. E., Matsumoto R., 2010, \apj, 712, 639

\bibitem[\protect\citeauthoryear{Oda \etal}{2012}]{Oda-etal12} 
Oda, H., Machida, M., Nakamura, K. E., Matsumoto, R. \& Narayan, R., 2012, \pasj, 64, 15

\bibitem[\protect\citeauthoryear{Paczy\'nski \& Wiita}{1980}]{Paczynski-Wiita80}
Paczy\'nski, B. and Wiita, P.J., 1980, \aap, 88, 23.

\bibitem[\protect\citeauthoryear{Peterson \etal}{2004}]{Peterson-etal04}
Peterson B. M., \etal, 2004, \apj, 613, 682

\bibitem[\protect\citeauthoryear{Radhika \& Nandi}{2014}]{Radhika-Nandi14}
Radhika D., Nandi A., 2014, AdSpR, 54, 1678

\bibitem[\protect\citeauthoryear{Riffel \etal}{2013}]{Riffel-etal13}
Riffel R. A., Storchi-Bergmann T., Winge C., 2013, MNRAS, 430, 2249

\bibitem[\protect\citeauthoryear{Samadi \etal}{2014}]{Samadi-etal14} 
Samadi, M., Abbassi, S. \& Khajavi, M., 2014, \mnras, 437, 3124 

\bibitem[\protect\citeauthoryear{Satyapal \etal}{2004}]{Satyapal-etal04}
Satyapal S., Sambruna R. M., Dudik R. P., 2004, \aap, 414, 825

\bibitem[\protect\citeauthoryear{Shafi \etal}{2015}]{Shafi-etal15}
Shafi N., Oosterloo T. A., Morganti R., Colafrancesco S., Booth
R., 2015, \mnras, 454, 1404 

\bibitem[\protect\citeauthoryear{Shakura \& Sunyaev}{1973}]{Shakura-Sunyaev73}
Shakura, N. I., Sunyaev, R. A., 1973, \aap, 24, 337S.

\bibitem[\protect\citeauthoryear{Shapiro \& Teukolsky}{1983}]{Shapiro-Teukolsky83} 
Shapiro S. L., Teukolsky S. A., 1983, Black Holes, White Dwarfs and 
Neutron Stars: The Physics of Compact Objects, A Wiley-Interscience Publication, New York.

\bibitem[\protect\citeauthoryear{Singh \& Chakrabarti}{2011}]{Singh-Chakrabarti11}
Singh C. B., Chakrabarti S. K., 2011, \mnras, 410, 2414

\bibitem[\protect\citeauthoryear{Sukov\'a \& Janiuk}{2015}]{Sukova-Janiuk15}
Sukov\'a P., Janiuk A., 2015, \mnras, 447, 1565

\bibitem[\protect\citeauthoryear{ Walsh \etal}{2013}]{Walsh-etal13}
Walsh J. L., Barth A. J., Ho L. C., Sarzi M., 2013, \apj, 770, 86

\bibitem[\protect\citeauthoryear{Yamauchi \etal}{2004}]{Yamauchi-etal04}
Yamauchi A., Nakai N., Sato N., Diamond P., 2004, \pasj, 56, 605

\bibitem[\protect\citeauthoryear{ Yuan \etal}{2002}]{Yuan-etal02}
Yuan F., Markoff S., Falcke H., 2002, \aap, 383, 854

\end{thebibliography}
\end{document}